\documentclass[10pt]{article}
\usepackage{fancyhdr}
\usepackage{extramarks}
\usepackage{amsmath}
\usepackage{amsthm}
\usepackage{amsfonts}
\usepackage{siunitx}
\usepackage{tikz}
\usepackage[plain]{algorithm}
\usepackage{algpseudocode}
\usepackage{multirow}
\usepackage{booktabs}
\usepackage{graphicx}
\usepackage{subfigure}
\usepackage[colorlinks,linkcolor=black,anchorcolor=black,citecolor=black,urlcolor=blue]{hyperref}
\usepackage{amsmath,bm}
\usepackage{booktabs}
\usepackage{mathtools}
\usepackage{amssymb}
\usepackage{caption}
\usepackage{capt-of}
\usepackage{mciteplus}
\usepackage{cite}
\usepackage{mathrsfs}
\usepackage[title,titletoc,toc]{appendix}
\usepackage{hyperref}
\usepackage[hyphenbreaks]{breakurl}
\usepackage{parskip}
\usepackage{soul}
\usepackage{textcomp}
\usepackage[colaction]{multicol}
\usepackage[switch]{lineno}
\usepackage{lipsum}
\usepackage{etoolbox}
\usepackage{longtable}
\usepackage{array}
\usepackage{tablefootnote}
\usepackage{ragged2e}
\newcolumntype{C}[1]{>{\centering\arraybackslash}p{#1}}
\usetikzlibrary{automata,positioning}
\usepackage{adjustbox}
\usepackage{makecell}
\usepackage{pdflscape}
\usepackage{setspace}
\usepackage{color}
\usepackage{authblk}

\usetikzlibrary{automata,positioning}
\usepackage{url}

\usepackage{xr}
\makeatletter
    \newcommand*{\addFileDependency}[1]{
    \typeout{(#1)}
    \@addtofilelist{#1}
    \IfFileExists{#1}{}{\typeout{No file #1.}}
    }

\makeatother
\newcommand*{\myexternaldocument}[1]{
    \externaldocument{#1}
    \addFileDependency{#1.tex}
    \addFileDependency{#1.aux}
}
\myexternaldocument{paper34_SI_clean}

\usepackage{url}

\begin{document}

\title{Machine Learning Study of the Extended Drug-target Interaction Network informed by Pain Related Voltage-Gated Sodium Channels }
\author[1]{Long Chen}
\author[1,2,*]{Jian Jiang}
\author[1]{Bozheng Dou}
\author[2]{Hongsong Feng}
\author[1]{Jie Liu}
\author[1]{Yueying Zhu}
\author[1]{Bengong Zhang}
\author[3]{Tianshou Zhou}
\author[2,4,5,+]{Guo-Wei Wei}

\affil[1]{Research Center of Nonlinear Science, School of Mathematical and Physical Sciences, Wuhan Textile University, Wuhan, 430200, P R. China}
\affil[2]{Department of Mathematics, Michigan State University, East Lansing, Michigan 48824, USA}
\affil[3]{Key Laboratory of Computational Mathematics, Guangdong Province, and School of Mathematics, Sun Yat-sen University, Guangzhou, 510006, P R. China}
\affil[4]{Department of Electrical and Computer Engineering Michigan State University, East Lansing, Michigan 48824, USA}
\affil[5]{Department of Biochemistry and Molecular Biology Michigan State University, East Lansing, Michigan 48824, USA}
\affil[*]{Corresponding author: jjiang@wtu.edu.cn}
\affil[+]{Corresponding author: weig@msu.edu}

\date{} 

\maketitle

\abstract{Pain is a significant global health issue, and the current treatment options for pain management have limitations in terms of effectiveness, side effects, and potential for addiction. There is a pressing need for improved pain treatments and the development of new drugs. Voltage-gated sodium channels, particularly Nav1.3, Nav1.7, Nav1.8, and Nav1.9, play a crucial role in neuronal excitability and are predominantly expressed in the peripheral nervous system. Targeting these channels may provide a means to treat pain while minimizing central and cardiac adverse effects.
In this study, we construct protein-protein interaction (PPI) networks based on pain-related sodium channels and develop a corresponding drug-target interaction (DTI) network to identify potential lead compounds for pain management. To ensure reliable machine learning predictions, we carefully select 111 inhibitor datasets from a pool of over 1,000 targets in the PPI network. We employ three distinct machine learning algorithms combined with advanced natural language processing (NLP)-based embeddings, specifically pre-trained transformer and autoencoder representations.
Through a systematic screening process, we evaluate the side effects and repurposing potential of over 150,000 drug candidates targeting Nav1.7 and Nav1.8 sodium channels. Additionally, we assess the ADMET (absorption, distribution, metabolism, excretion, and toxicity) properties of these candidates to identify leads with near-optimal characteristics. Our strategy provides an innovative platform for the pharmacological development of pain treatments, offering the potential for improved efficacy and reduced side effects.
}

\maketitle

\newpage

\section{Introduction} 
Pain is a complex phenomenon and can be categorized in various ways based on different factors, such as acute pain and chronic pain, nociceptive pain and neuropathic pain, etc. Pain, with distinct types, has been estimated to occur in probably 35\% of the population in the United States, with a higher morbidity rate than cancer and heart disease \cite{news}. Pain management is a branch of medicine that utilizes an interdisciplinary approach. Although intensive efforts have been made to design new drugs for pain management over the last decades, almost half of patients with chronic pain show little response to existing analgesic drugs. Hence, there is an urgent need to design new drugs for pain treatment.

Voltage-gated sodium channels (Na$\text{v}$s) are integral membrane proteins that play a crucial role in the generation and propagation of action potentials in neurons and other excitable cells. These channels are responsible for the rapid influx of sodium ions into the cell, which leads to depolarization and the initiation of an action potential. More specifically, Na$\text{v}$s modulate membrane permeability to sodium ions and facilitate important intercellular functions, which are related to a variety of diseases, including chronic pain, cardiac arrhythmia, and others. Particularly, Na$\text{v}$s subtypes, such as Na$\text{v}$1.3, Na$\text{v}$1.7, Na$\text{v}$1.8, and Na$\text{v}$1.9, encoded respectively by the genes SCN3A, SCN9A, SCN10A, and SCN11A, present the best opportunities for pain therapeutics, since they almost exclusively distribute in the peripheral nervous system and highly express in sympathetic ganglia, olfactory epithelium, and dorsal root ganglion sensory neurons \cite{dib2013nav1}. Na$\text{v}$1.3, originally termed sodium channel III, was cloned and sequenced in the 1980s from rat brain tissue \cite{noda1986existence}. In 2006, it was validated to be critical to pain transmission and modulation pathways \cite{RIECHERS2015567}.

The expression level of Na$\text{v}$1.7 was found to be related to pain based on animal models. Specifically, gain-of-function and loss-of-function mutations in humans result in extreme pain disorder and insensitivity to pain, respectively \cite{black2004changes,fertleman2006scn9a,cox2006scn9a}. This suggests that Na$\text{v}$1.7 plays a vital role in pain generation, making it a hot target for pain treatment in recent years.
Na$\text{v}$1.8 is preferentially expressed in peripheral sensory neurons. It has been shown to shape action potentials in these neurons and contribute to pain phenotypes in humans and animal studies. The key role of Na$\text{v}$1.8 in repetitive firing, and its localization in free nerve endings where the response to external stimuli is integrated and action potentials are initiated, indicates that Na$_\text{v}$1.8 can play a strong part in nociception and chronic pain \cite{black2008multiple,rowe2013voltage}.
Additionally, recent genetic and functional findings linking Na$\text{v}$1.9 to human pain disorders have suggested that Na$\text{v}$1.9 is a vital contributor to pain in humans, including its pattern of expression, subcellular localization, and modulation \cite{okuda2016infantile,leipold2015cold,han2017familial,han2015domain}.
Subsequently, many kinds of Na$\text{v}$1.3, Na$\text{v}$1.7, Na$\text{v}$1.8, and Na$\text{v}$1.9 inhibitors have been found for pain treatment, including sulfonamides, guanidium compounds, and cystine knot peptides \cite{mulcahy2019challenges}. However, the specific roles of these pain-related Na$_\text{v}$s in the generation and transmission of pain signals remain unknown and are still being actively researched.

It is well-known that proteins do not function independently in cells and organisms. Protein-protein interactions (PPIs) play a fundamental role in virtually all biological processes, including DNA replication, transcription, translation, protein folding, intracellular signaling, and metabolism. Therefore, it is crucial to understand the role of Na$\text{v}$-inferred PPI networks in pain generation, management, treatment, and therapeutic development. Na$\text{v}$-inferred PPI networks can be used to systematically analyze potential treatment efficacy and side effects. The nodes of a PPI network represent proteins, and the links or edges represent direct or indirect interactions between nodes that contribute to certain biological activities. The String Database v11 (https://string-db.org/) can be utilized to build a PPI network as it provides a large collection of protein-protein interactions for given proteins or diseases. In the study of sodium channels, we can build the PPI networks related to the major sodium channels involved in pain, such as Na$\text{v}$1.3, Na$\text{v}$1.7, Na$\text{v}$1.8, and Na$\text{v}$1.9, based on which we can carry out systematic analysis of medication treatment and side effects.

The proteins in these PPI networks are the test targets for treatment or side effects. However, traditional in vivo or in vitro assay tests are highly time-consuming. High-throughput screening in experiments has been employed to find these inhibitors, but it is time-consuming and resource-intensive, making it unsuitable for screening a large collection of drug candidates in drug discovery. Moreover, large-scale experiments on animals raise legal and ethical issues. Hence, Artificial Intelligence (AI), including machine learning (ML) methods, can be employed in this study for large-scale predictions \cite{bagherian2021machine}.
 
Artificial intelligence drug design (AIDD) has been considered capable of providing accurate computational predictions and speeding up drug development. It offers low costs and the ability to find optimally structured compounds with the help of ML algorithms and large availability of experimental data. Recently, many advanced ML methods have been applied to pain treatment.  Lomartire et al. analyzed the data of a large population-representative sample of chronic pain patients and identified future sickness absence that should be considered when adapting interdisciplinary treatment programs to the patient's needs\cite{lomartire2021predictors}. Using machine learning, Miettinen et al. show that sleep as a core factor in chronic pain 
\cite{miettinen2021machine}.
Machine learning and  multiplex in situ hybridization were used to assign transcriptomic class in the trigeminal ganglion \cite{von2020assigning}.
 Additionally, Robinson et al. used machine classification algorithms to measure the difference between neuroimaging data and self-report in their ability to classify  individuals with and without chronic pain\cite{robinson2015comparison}.

Currently, numerous in silico methods have been developed for virtual screening of sodium channel inhibitors. Molecular fingerprint-based characterization of molecules is particularly popular, and classification studies on specific targets often yield good performance based on ligand structure and properties \cite{avram2018modeling,bagherian2021machine}.
For example, protein-ligand binding models for hERG (human ether-a-go-go potassium channel) have been proposed, achieving good classification results on hERG blockage using the Online Chemical Modeling Environment (OCHEM) \cite{li2017modeling, zhang2022hergspred, feng2023virtual}. Kong et al. \cite{kong2020prediction} developed a molecular group optimization method by combining the Grammar Variational Autoencoder, a classification model, and simulated annealing to predict Na$\text{v}$1.7 sodium channel inhibitors. They found that the random forest algorithm with CDK fingerprint performs best in imbalanced data sets. They also employed multiple ML methods to predict Na$\text{v}$1.5 inhibitors \cite{kong2023multiple}. Bosselmann et al. \cite{bosselmann2022learning} built a multi-task multi-kernel learning framework to improve the prediction of functional effects of missense variants in voltage-gated sodium channels based on phenotypic similarity. Additionally, Herrera et al. \cite{herrera2022pep} developed a bioinformatics tool called PEP-PRED$^{\text{Na+}}$ for highly specific prediction of voltage-gated sodium channel blocking peptides. This tool is helpful in accelerating and reducing the costs of designing new sodium channel blocking peptides with therapeutic potential.

More studies on voltage-gated sodium channels can be found in review papers \cite{nguyen2022towards,jenssen2021machine,matsangidou2021machine,lotsch2018machine}, which describe recent progress and future opportunities for developing sodium channel-targeting small molecules and peptides as non-addictive therapeutics to treat pain. However, these studies primarily focus on individual sodium channels and lack consideration of drug-target interaction networks, as well as comprehensive ADMET (absorption, distribution, metabolism, excretion, and toxicity) analysis.

Pain management is not limited to sodium channels and related inhibitors. Opioids, also known as narcotics, have been used for centuries in the treatment of pain. The use of opioids arose partially from the need to treat severe injuries sustained in warfare. Opioids can bind to opioid receptors, such as mu, kappa, and delta, on nerve cells in the brain, spinal cord, and other parts of the human body, blocking pain messages from reaching the brain, either from the body or the spinal cord. However, evidence suggests that long-term opioid use carries an increased risk of opioid use disorder (OUD) and opioid overdose, as well as various other adverse side effects, such as sleepiness, constipation, and nausea \cite{brady2016prescription}. The risk of addiction is particularly high when opioids are used for long periods to manage chronic pain. Recently, more attention has been focused on the treatment of OUD and drug addiction issues.

Feng et al. constructed an extended drug-target interaction (DTI) network based on the four major opioid receptors \cite{feng2023machine2}. They developed advanced machine learning predictors to study the screening and repurposing potential of tens of thousands of compounds in the opioid DTI network for OUD management. The results of this work were used to analyze the repurposing potential of thousands of DrugBank compounds and evaluate their ADMET properties \cite{feng2023machine}. Additionally, Zhu et al. built a topology-inferred drug addiction learning (TIDAL) model to analyze the opioid DTI network and address the problem of drug addiction \cite{zhu2023tidal}. Although opioid-based medications have been successfully used to treat acute postsurgical and postprocedural pain, their high risks and side effects have raised significant concerns. There is a need for safer and more effective drugs for pain treatment.

\begin{figure}[!tpb]
\centering
\includegraphics[width=15cm]{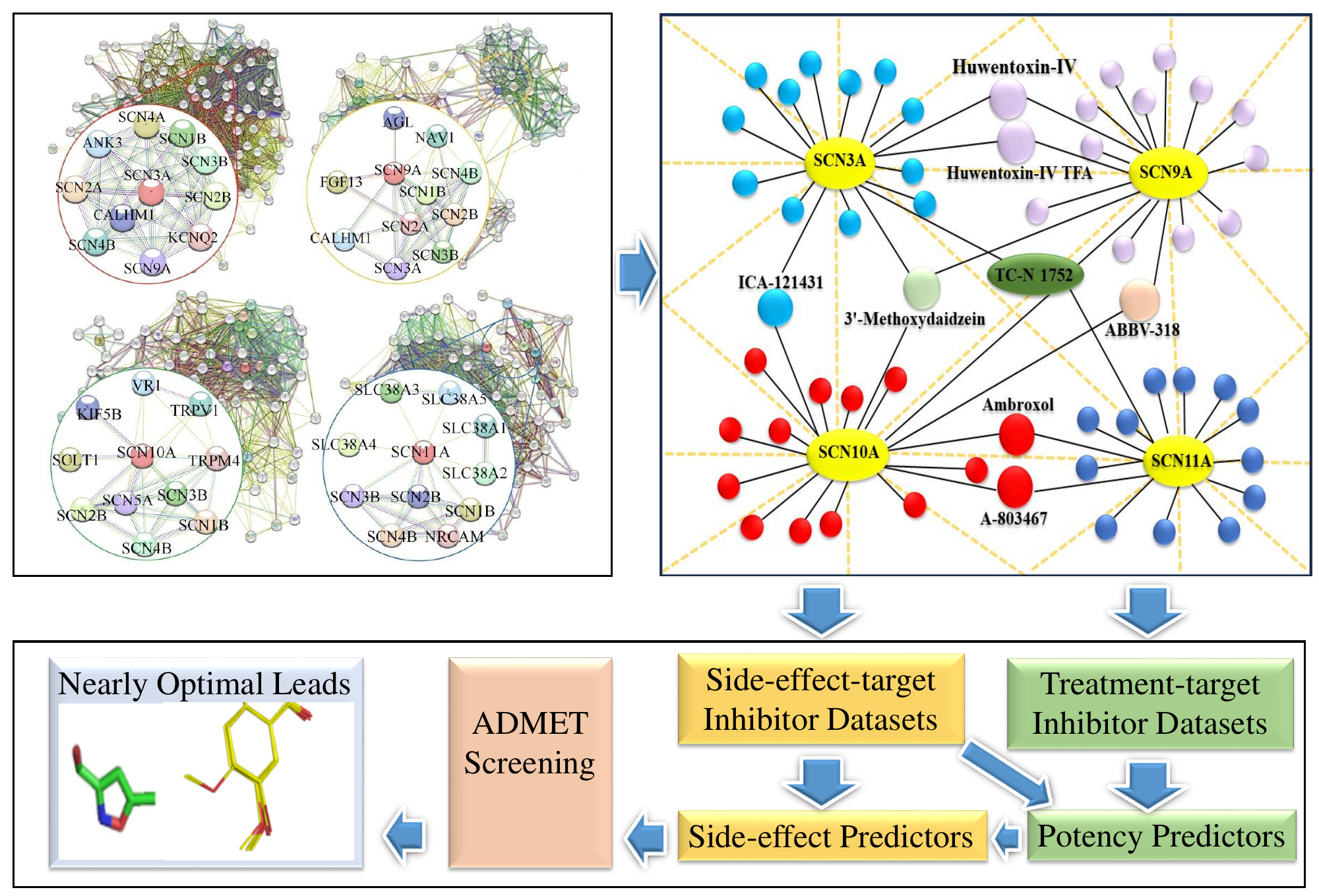}
\caption{The flowchart of screening nearly optimal lead compounds for inhibiting pain related voltage-gated sodium channels (VGSCs). 
Top left chart:  Protein-protein interaction (PPI) networks of the four VGSCs involve over 1,000 proteins, including SCN3A, SCN9A, SCN10A, and SCN11A. Each of them has a core and global PPI network. 
Further details of the PPI networks are provided in the Table S1 in the Supporting Information. 
Top right chart: The drug-target interaction (DTI) network involves 111 targets and 15,047 inhibitor compounds. Here only four targets (SCN3A, SCN9A, SCN10A, and SCN11A) with several compounds are displayed for simplicity. The yellow dashed lines indicate the connections among 111 targets. 
Low chart: predictive models for side effect and repurposing evaluation, as well as 
ADMET screening. }
\label{flowchart}
\end{figure}

In the present work, we construct an extended drug-target interaction (DTI) network informed by pain-related voltage-gated sodium channels, namely Na$\text{v}$1.3, Na$\text{v}$1.7, Na$\text{v}$1.8, and Na$\text{v}$1.9. We develop advanced machine learning (ML) models using natural language processing (NLP) tools, such as autoencoders and transformers, to study this DTI network.
Firstly, we build protein-protein interaction (PPI) networks of the four pain-related sodium channels from the String Database v11. This results in hundreds of related proteins that are considered potential side effect targets in our study. Secondly, we collect inhibitor datasets with experimental binding affinity labels from the CHEMBL database for these PPI targets, creating an extended DTI network with hundreds of targets and hundreds of thousands of drug candidates.
Thirdly, to generate our ML models, we embed the inhibitor compounds using two NLP models: a transformer and an autoencoder. The resulting latent feature vectors are combined with gradient boosting decision tree (GBDT), support vector machine (SVM), and random forest (RF) algorithms to build binding affinity (BA) prediction models.
Fourthly, we perform cross-predictions to screen side effects and repurposing potentials of over 150,000 compounds. Through these models, we evaluate the side effects of FDA-approved drugs or other existing medications and search for promising lead compounds.
Finally, in order to identify lead compounds, we also assess the pharmacokinetic properties in compound filtering, such as absorption, distribution, metabolism, excretion, toxicity (ADMET), and synthesizability. These steps are illustrated in Fig.\ref{flowchart}. Our study of the extended DTI network provides an innovative strategy for analyzing pain management and developing therapeutics.

\section{Results}

\subsection{Pain related voltage-gated sodium channel informed drug-target interaction (DTI) networks}

Voltage-gated sodium channels (VGSCs), which consist of a family of nine distinct proteins or genes (Na$\text{v}$1.1-1.9), exhibit different pharmacological properties. Specifically, the proteins Na$\text{v}$1.3, Na$\text{v}$1.7, Na$\text{v}$1.8, and Na$_\text{v}$1.9 are involved in neuropathic pain and are associated with both human Mendelian pain disorders and common pain disorders such as small-fiber neuropathy \cite{bennett2019role}. These four VGSC proteins play a role in modulating different types of pain, offering potential for the development of specific sodium channel inhibiting agents for chronic pain treatment.
Functionally, Na$\text{v}$1.7 is classified as tetrodotoxin-sensitive (TTX-S), while Na$\text{v}$1.8 and Na$\text{v}$1.9 are considered tetrodotoxin-resistant (TTX-R). Anatomically, these proteins exhibit broad and distinct expression patterns across neuronal and smooth muscle cells throughout the body, as well as in cells of the immune system where they participate in migration and phagocytosis \cite{erickson2018voltage}. Traditionally, Na$\text{v}$1.3 is primarily expressed in the brain and spinal cord, while Na$\text{v}$1.7, Na$\text{v}$1.8, and Na$_\text{v}$1.9 tend to be expressed in the peripheral nervous system.
Furthermore, these channels are regulated by a variety of enzymes and structural proteins, such as kinases, auxiliary $\beta$-subunits, and ubiquitin-protein ligases, which collectively influence sodium channel biophysical properties and expression \cite{tseng2007sodium,laedermann2015post}.

Pain-related VGSCs are widely distributed throughout the body, and their interactions with various upstream and downstream proteins play a crucial role in specific biological functions. To analyze these interactions, we constructed protein-protein interaction (PPI) networks centered around each of the four pain-related VGSCs. The gene names SCN3A, SCN9A, SCN10A, and SCN11A were used as inputs to the String database to extract the corresponding PPI networks. The resulting networks, shown in the top left panel of Fig.\ref{flowchart}, represent direct and indirect interactions between proteins and each pain-related VGSC. Each PPI network contains 401 proteins, focusing on critical interactions rather than considering a larger number of proteins. It is important to note that there is some overlap between the networks, indicating interdependencies among the VGSCs.

Considering that compounds that act as agonists or antagonists on pain-related VGSCs can influence their pharmacological behavior in pain treatment, we aimed to identify additional compounds that bind to these VGSCs. To evaluate the binding effects of inhibitors on VGSCs and other proteins in the PPI networks, we searched and collected inhibitor compounds from the Chembl database for each protein. This process resulted in an extended drug-target interaction (DTI) network, encompassing 111 targets or related datasets and a total of 150,147 inhibitor compounds, which is illustrated in the top right panel of Fig.\ref{flowchart}. The protein names of these 111 datasets are listed in Table S2 in the Supporting Information, and additional details about the collected datasets can be found in Table S3 in the Supporting Information.

\subsection{Binding affinity predictions for the extended DTI network}

\begin{figure}[!tpb]
\centering
\includegraphics[width=16cm]{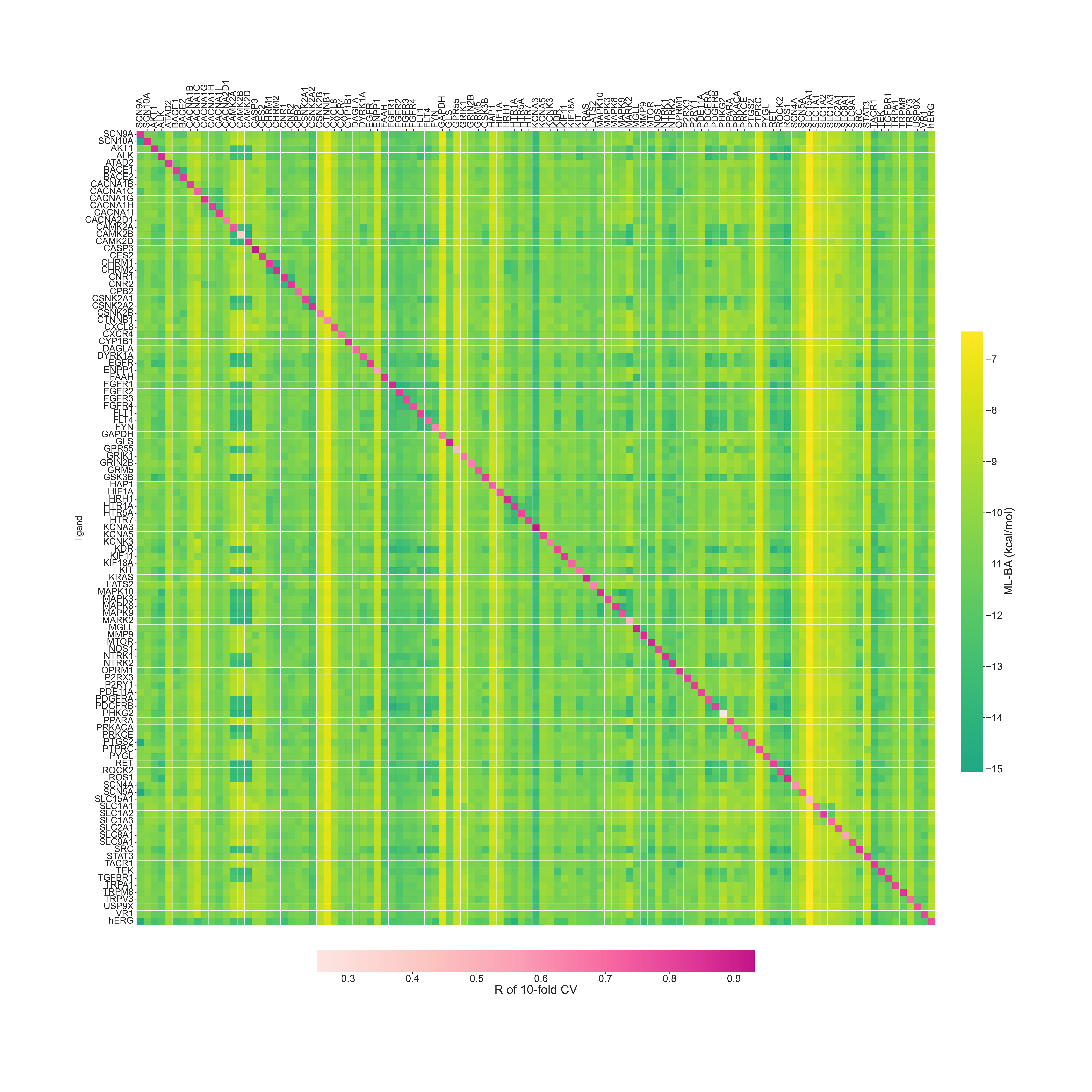}
\caption{The heatmap of cross-target binding affinities (BAs) predictions for the extended DTI networked informed by four pain-related voltage-gated sodium channels. The left labels of the heatmap represent all the inhibitor datasets and   those above the heatmap mean the machine learning (ML) models. The diagonal elements in the heatmap denote the Pearson correlation efficient ($R$) of ten-fold cross validation for all the ML models. The off-diagonal elements in each row indicate the highest BAs values of inhibitors of one datasets predicted by 111 ML models. This heatmap is used to reveal the inhibitor specificity of each dataset on other protein targets.}
\label{heatmap}
\end{figure}

Using autoencoder and transformer embeddings, we developed 111  ML  models for all 111 targets and 150,147 compounds in the extended  DTI network. The cross-target binding affinity (BA) predictions were carried out using these 111 ML models, and the results are presented in Fig.\ref{heatmap}. The diagonal elements of the heatmap represent the Pearson correlation coefficient ($R$) obtained from ten-fold cross-validation for each ML model. The mean, maximum, and minimum values of $R$ across the models are 0.77, 0.93, and 0.25, respectively. Notably, 53 models achieved $R$ values greater than 0.8, indicating high predictive performance.

Furthermore, the root mean square error (RMSE) values of these models, as shown in Table S3 in the Supporting Information, range from 0.43 to 1.15 kcal/mol. These values fall within a reasonable range, suggesting that the ML models exhibit excellent prediction accuracy and reliable performance for binding affinity predictions.

\subsubsection{Cross-target binding affinity predictions for the extended DTI network}

In this section, we conduct an analysis of compound cross-target interactions to estimate their side effects on other proteins in the protein-protein interaction (PPI) network, providing a better understanding of the extended drug-target interaction (DTI) network. The off-diagonal elements of the heatmap in Fig.\ref{heatmap} represent the maximum binding affinity (BA) values (i.e., BA with the largest absolute values) of inhibitor compounds from one dataset predicted by other ML models. The labels on the left side of the heatmap correspond to the 111 inhibitor datasets, while the labels on the top of the heatmap correspond to all the 111 ML models. Each column in the heatmap represents the predictions made by a specific model.

For instance, the $i$-th element in the $j$-th column indicates the prediction result of the $i$-th dataset by the $j$-th model. These cross-target prediction results serve as indicators of the potential side effects of one inhibitor dataset on other proteins. In our analysis, we use an inhibition threshold value of -9.54 kcal/mol ($K_i=0.1 \mu M$) for the BA values\cite{flower2002drug}. If a compound has a BA value below this threshold, it is considered active in terms of its biological function. Otherwise, it is classified as an inactive compound.

According to our analysis, out of the 12,210 cross-predictions, 9,262 were found to exhibit side effects based on this threshold value, as their predicted maximal BA values were below -9.54 kcal/mol. Additionally, the remaining 2,948 cross-prediction results showed weak side effects, as their maximal BA values exceeded -9.54 kcal/mol. The color of the off-diagonal elements in the heatmap indicates the strength of the side effects, with closer proximity to green representing stronger side effects, and closer proximity to yellow indicating weaker side effects.

It is worth noting that in Fig.\ref{heatmap}, several yellow vertical lines can be observed, suggesting very slight predicted side effects on these proteins. This could be due to the majority of collected experimental BA labels being larger than -9.54 kcal/mol, which limits the predictive power of the ML models in such cases.

The reasons for side effects caused by drug candidates targeting a specific protein are often complex, and one possible factor is the presence of similar binding sites on off-target proteins. Proteins within the same family often share similar structures or sequences, leading to the existence of comparable binding sites. As a result, an inhibitor compound that is effective against one protein may also bind to another protein within the same family, giving rise to mutual side effects.

As observed in Fig.\ref{heatmap}, mutual side effects occur among the three targets CAMK2A, CAMK2B, and CAMK2D, which belong to the calmodulin-dependent protein kinase II (CAMK2) family and share similar 3D structural conformations or 2D sequences. This observation is further supported by the alignments of their 3D structures and 2D sequences, as shown in Fig. S1 of the Supporting Information.

We can identify more examples of mutual side effects among proteins within the same family. For instance, the fibroblast growth factor target (FGFR) family, which includes FGFR1, FGFR2, FGFR3, and FGFR4, as well as the mitogen-activated protein kinase (MARK) family, which comprises MARK2, MARK3, MARK8, MARK9, and MARK10, exhibit mutual side effects. These examples illustrate the occurrence of mutual side effects among proteins in the same family, emphasizing the importance of considering family-wide effects in drug development and analysis.

\subsubsection{Predictions of side effects and repurposing potentials for the extended DTI network}

\begin{figure}[!tpb]
\centering
\includegraphics[width=14cm]{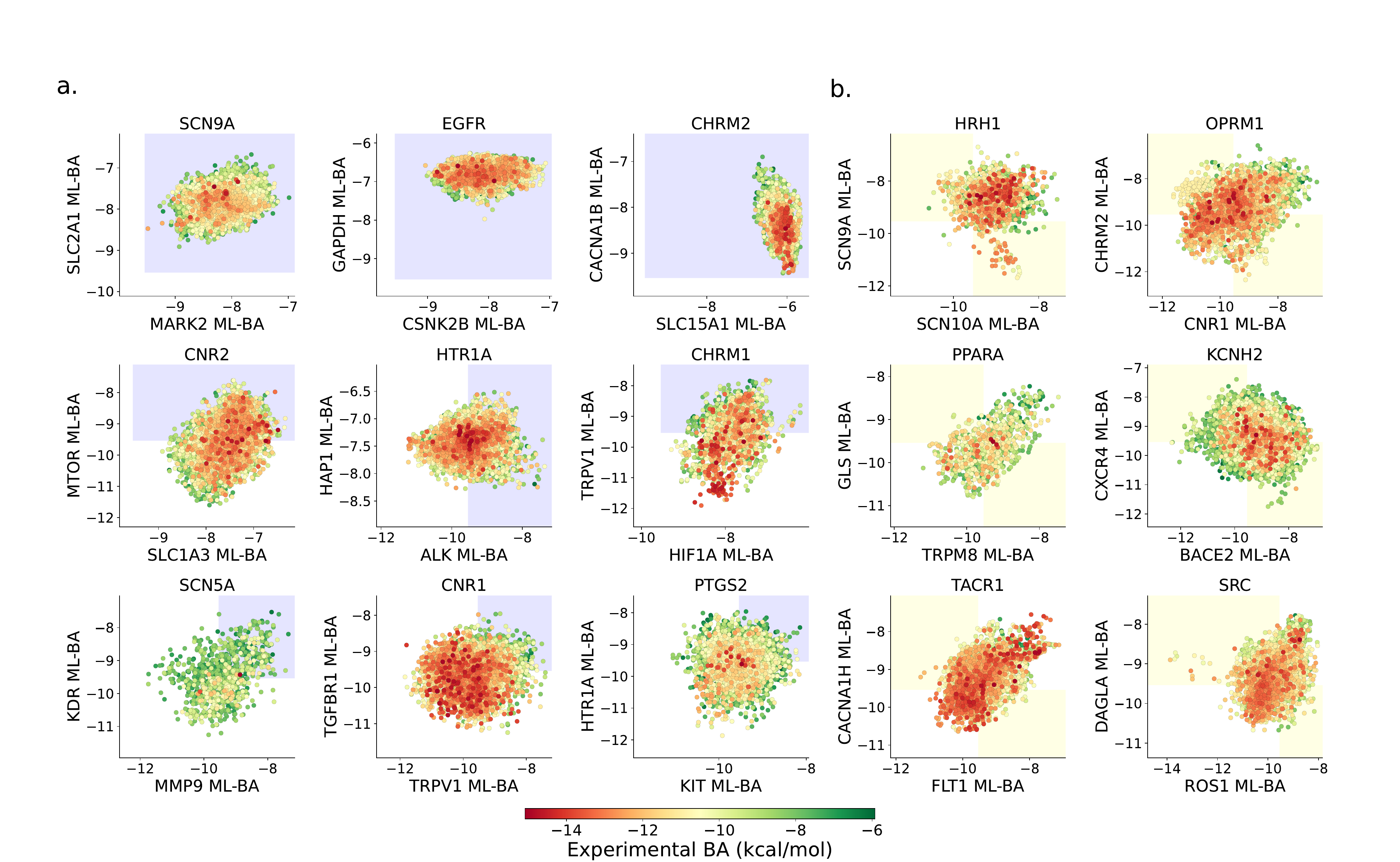}
\caption{Examples of predictions of side effects and repurposing potentials. 
$\bf a$ 
The first row, second row, and third row represent example inhibitor datasets that have side effects on none, one, and two of the given two off-target proteins, respectively.  The blue frames indicate where there is no side effects. 
$\bf b$ The first row, second row, and third row represent example inhibitor datasets that are equipped with repurposing potentials on none, one, and two of the two off-target proteins. The yellow frames indicate the inhibitors have repurposing potential for one protein but have no side effect on the other protein.  }
\label{repurposing}
\end{figure}

Side effects occur when a drug candidate exhibits strong binding affinity to the intended target but inadvertently affects other proteins as potential off-target inhibitors. These side effects can be identified through cross-target predictions, as illustrated in Fig.\ref{repurposing}a, for the extended DTI network. Each panel in the figure represents a specific target protein and two corresponding off-target proteins, indicated by the panel title, $x$-axis, and $y$-axis, respectively. The scattered points in the plot are color-coded based on the experimental binding affinities (BAs) of the inhibitors for the target protein. Red and green colors represent high and low binding affinities, respectively. The $x$-axis and $y$-axis values represent the predicted BAs obtained from two machine learning (ML) models constructed using inhibitor datasets for the two off-target proteins.

The blue frames in the nine panels of Fig.\ref{repurposing}a indicate regions where no side effects are predicted on the two off-target proteins. The three rows of the figure represent different scenarios for inhibitors targeting a specific protein, showing the presence of side effects on zero, one, or both of the given off-target proteins.
For instance, in the first panel of the first row, all inhibitors for protein SCN9A are predicted to have weak inhibitory effects, with binding affinity (BA) values greater than -9.54 kcal/mol, on the two off-target proteins. In the first panel of the second row, approximately half of the inhibitors for protein CNR2 are predicted to exhibit strong binding affinity to the MTOR protein, while none of the inhibitors are predicted to bind to the SLC1A3 protein. Furthermore, in the second panel of the third row, most inhibitors of protein CNR1 are predicted to efficiently bind to both the TGFBR1 and TRPV1 proteins simultaneously.

The repurposing potential of inhibitors can also be determined through cross-target predictions. Drug candidates that exhibit weak binding affinity to their designated targets but potent inhibition of other proteins are defined to possess repurposing potential. Fig.\ref{repurposing}b displays six prediction cases of repurposing identified by our models. In the yellow frames, the inhibitors for the target protein exhibit strong binding to one protein (i.e., predicted BAs less than -9.54 kcal/mol), but weak binding to the other protein (i.e., predicted BAs greater than -9.54 kcal/mol).
For example, in the first panel of the first row in Fig.\ref{repurposing}b, many inhibitors for protein HRH1 are predicted to have repurposing potential for either SCN9A or SCN10A, but not for the other one. Since both SCN9A and SCN10A are important targets for drug design in pain treatment, it is crucial to identify more drug candidates for these two proteins through the virtual screening process. Carbamazepine, a voltage-dependent Nav1.7 sodium channel (SCN9A) blocker, has undergone a phase I clinical study in humans \cite{mann2019review}. Our models can be employed to find more inhibitors that can bind to SCN9A, similar to the mechanism of Carbamazepine. The second and third rows in Fig.\ref{repurposing} depict additional cases where inhibitors for a given protein have repurposing potential for two other proteins.

\subsubsection{Protein similarity inferred by cross-target  correlations in the DTI network}

\begin{figure}[!tpb]
\centering
\includegraphics[width=14cm]{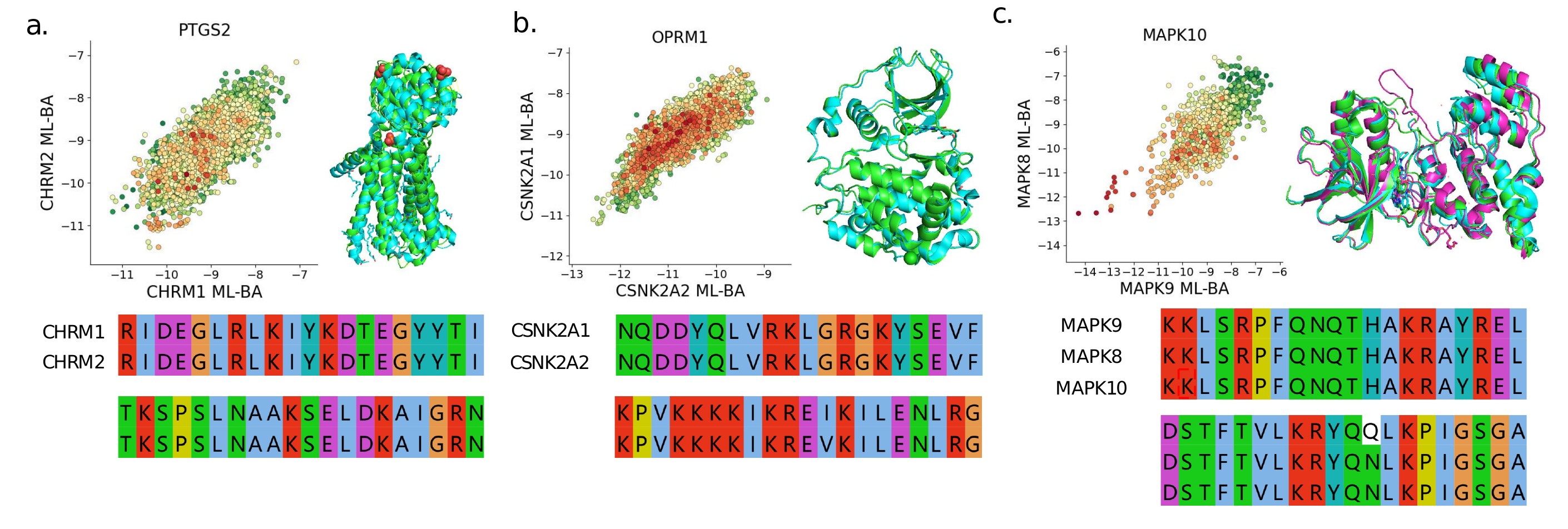}
\caption{Three examples of correlated predicted BA values suggesting the structure and/or sequence similarities of proteins. In each panel, the $x$-axis and $y$-axis represent the predicted BA values on two other proteins, and the scattered points with colors indicate the experimental labels of inhibitors of the target. The 3D structure alignment is shown in the right of the panel, and the 2D sequence alignment is shown below. In the 3D structure alignment, PDB 6ZG4 and 3UON are used for CHRM1 and CRMH2, PDB 6QY7 and 6QY9 for CSNK2A1 and CSNK2A2, PDB 3ELJ, 7N8T, and 3KVX for MAPK8, MAPK9, and MARK10, respectively. }
\label{alignment}
\end{figure}

As side effects can arise when a drug candidate binds to proteins with similar 3D structures or sequences, the predicted BA values in cross-target BA prediction may exhibit correlation. In other words, correlated predicted BA values can serve as an indication of similar binding sites or 3D protein structures. Fig.\ref{alignment}a illustrates a linear correlation between the predicted BAs of inhibitors for PTGS2 on CHRM1 and CHRM2 proteins, with a Pearson correlation coefficient $R$ of up to 0.71. The high correlation is attributed to the high binding site similarity between CHRM1 and CHRM2 proteins, as validated by the alignments of 3D structures and 2D sequences in Fig.\ref{alignment}a. The 3D structures of the two proteins were found to be quite similar, and the identity of the 2D binding site sequence reached as high as 63\%.

Two additional examples can be observed in Fig.\ref{alignment}b and c, demonstrating that the predicted BA correlation indicates similar 3D protein structures. The Pearson correlation coefficients are 0.82 and 0.72 for the cases in Fig.\ref{alignment}b, corresponding to the predicted BAs for OPRM1 on MARK9 and MARK8, respectively. These alignments of 3D structures and 2D sequences validate the usefulness of cross-prediction in detecting protein similarity.

Furthermore, Fig.\ref{alignment}c reveals a bilinear correlation relationship, where the predicted BAs of MAPK9 inhibitors not only linearly correlate with MARK8 and MARK10 proteins, but also exhibit a linear correlation with their experimental BA values, as indicated by the color coding. This bilinear relationship is confirmed by the alignment of 3D structures and 2D sequences of the three proteins. This result suggests that a potent MAPK9 inhibitor is likely to be a strong binder for both MARK8 and MARK10 proteins simultaneously. The high structural similarities result in a drug-mediated trilinear target relationship. The observed bilinear or trilinear relationship indicates the possibility of developing inhibitors that can bind to multiple targets of major pain proteins simultaneously.

\subsection{Druggable property screening}

Evaluation of ADMET   is of utmost importance in drug design and discovery. ADMET encompasses several essential attributes that are correlated with the pharmacokinetic study of a compound. A promising drug candidate should not only exhibit potency against the therapeutic target but should also possess favorable ADMET properties.
Furthermore, hERG   is a crucial potassium ion channel known for its contribution to the electrical activity of the heart. When this channel is blocked by a drug, it can lead to serious side effects on the heart. Therefore, the evaluation of hERG risk is indispensable in drug development and assessment.

In this section, we conducted the evaluation of ADMET using six indexes, namely FDAMDD, T$_{1/2}$, F$_{20\%}$, logP, logS, and Caco-2, along with synthetic accessibility (SAS) and hERG risk assessment.
FDAMDD represents the FDA maximum recommended daily dose, which aims to avoid toxicity in the human body. The half-life (T$_{1/2}$) refers to the time it takes for the concentration of a drug in the body to decrease by half. A value of T$_{1/2}$ less than three hours indicates a shorter half-life.
F$_{20\%}$ represents the probability of an administered drug reaching systemic circulation with less than 20\% of the initial dose. This parameter is important for assessing the effectiveness, bioavailability, therapeutic efficacy, and potential side effects of a drug.
LogP refers to the logarithm of the partition coefficient of a compound between a nonpolar solvent and water, providing information about its hydrophobicity. On the other hand, logS represents the logarithm of the aqueous solubility of a compound, which indicates its ability to dissolve in water.
Caco-2 is a measure used to estimate the in vivo permeability of oral drugs. It provides valuable information about a drug candidate's interaction with efflux transporters, metabolism, and other factors that influence its absorption.
SAS is employed to assess the feasibility of synthesizing a specific compound or molecule, taking into account its structural complexity and the availability of synthetic routes.

\begin{figure}[!tpb]
\centering
\includegraphics[width=16cm]{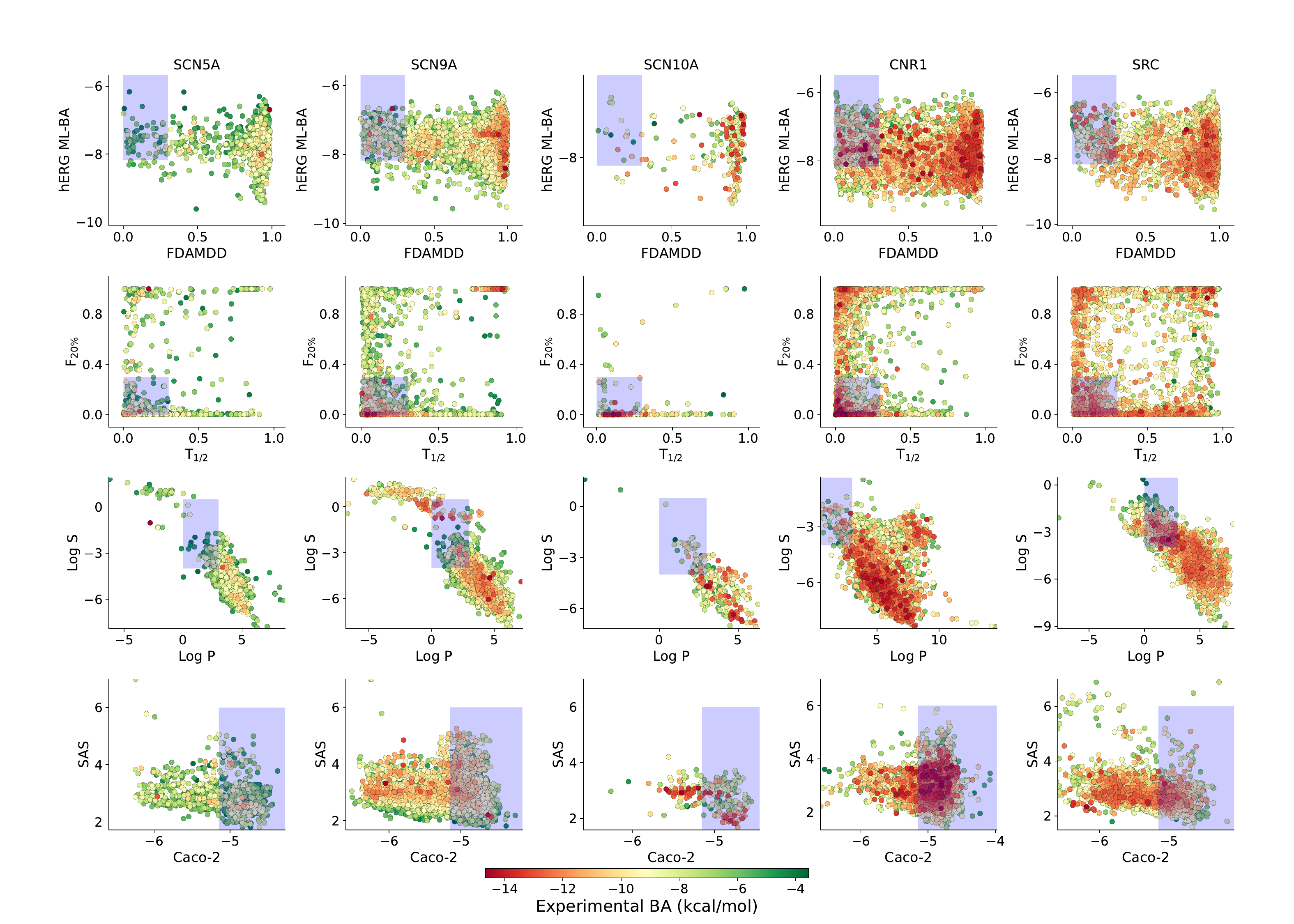}
\caption{Druggable property screening based on ADMET properties, synthesizability, and hERG side effects on compounds from five protein datasets: SCN5A, SCN9A, SCN10A, CNR1, and SRC. The colors of the points indicate the experimental BAs for these targets. The $x$- and $y$-axes represent various predicted ADMET properties, synthesizability, or hERG side effects. Blue frames highlight the optimal ranges of these properties and side effects.}
\label{ADMET}
\end{figure}

During the above estimation in the present work, ADMETlab 2.0 (\url{https://admetmesh.scbdd.com/}) solvers were used for ML predictions and provided a set of optimal ranges for these ADMET properties \cite{xiong2021admetlab}. The SAS assessment was implemented using Rdkit packages \cite{landrum2013rdkit}. The optimal ranges of ADMET properties and SAS are listed in Table \ref{table_1}, in which a stricter threshold of -8.18 kcal/mol ($K_i = 1 \mu M$) is applied to exempt hERG side effects. Fig.\ref{ADMET} illustrates the ADMET screening of five inhibitor datasets, including SCN5A, SCN9A, SCN10A, CNR1, and SRC, that play essential roles in pain treatment. The first row of Fig.\ref{ADMET} depicts the distributions of FDAMDD and hERG side effects of inhibitors from the five datasets. The blue frames represent the optimal domains of the two properties mentioned above. The colors of the points indicate the experimental BA values for targets. From this screening, all five datasets have sufficient compounds with optimal toxicity and hERG side effects. However, for the SCN10A dataset, there are only a few potent inhibitors in the optimal domains. This suggests that ADMET properties and side effects should be taken into account before synthesizing a new compound.

The second row of Fig.\ref{ADMET} displays the screening results on absorption properties: T$_{1/2}$ (half-life) and F$_{20\%}$ (bioavailability 20\%). It is observed that for all five datasets, the optimal domain of T$_{1/2}$ and F$_{20\%}$ occupies only a small fraction of chemical space. This indicates a strict screening process, emphasizing the critical roles of these two properties in physicochemical assessment.

The third row of Fig.\ref{ADMET} illustrates the screening for logP and logS, which are closely related to the distribution of chemicals in the human body. In all five datasets, only a small portion of potent inhibitors is found within the optimal domain, suggesting that a large number of inhibitors are not well absorbed in the human body.

The last row of Fig.\ref{ADMET} presents the screening results for Caco-2 and SAS. These five plots demonstrate that almost all compounds from the five datasets are easy to synthesize, and approximately half of the compounds exhibit good cell permeability. Notably, a significant number of potent inhibitors fall within the optimal domain.

\begin{table*}[h]\footnotesize
	\centering
     \caption{The optimal ranges of selected ADMET properties and synthetic accessibility (SAS) used for screening compounds in this work.}\label{table_1}
      \begin{tabular}{cc}
        \hline
        Property & Optimal ranges\\
        \hline
        FDAMDD & Excellent: 0–0.3; medium: 0.3–0.7; poor: 0.7–1.0\\
         F$_{20\%}$ & Excellent: 0–0.3; medium: 0.3–0.7; poor: 0.7–1.0\\
         Log P & The proper range: 0–3 log mol/L \\
         Log S & The proper range: -4-0.5 log mol/L\\
         T$_{1/2}$ & Excellent: 0–0.3; medium: 0.3–0.7; poor: 0.7–1.0 \\
         Caco-2 & The proper range: $> -5.15$\\
         SAS & The proper range: $< 6$  \\
        \hline
    \end{tabular}
\end{table*}

\section{Discussion}
	
\subsection{Side effect evaluations of existing medications for pain treatment}

SCN3A, SCN9A, SCN10A, and SCN11A are genes that encode sodium channels in the Navs family. These channels play an important role in the generation and propagation of action potentials in neurons, including those involved in pain signaling. Additionally, it has been found that blocking these channels could reduce pain hypersensitivity. There are several FDA-approved experimental medications available for the treatment of pain, which can be roughly classified into four classes: non-opioid analgesics, nonsteroidal anti-inflammatory drugs (NSAIDs), opioid medications, and others. In this study, we utilized our DTI-based ML models to predict the side effects of these medications.
   
Acetaminophen, commonly known as Tylenol or paracetamol, is a typical over-the-counter non-opioid analgesic used to temporarily relieve mild to moderate pain, such as headaches, muscular aches, backaches, toothaches, and premenstrual and menstrual cramps. It is a weak inhibitor of both cyclooxygenase (COX)-1 and COX-2 in vitro and eases pain by inhibiting the production of prostaglandins, which are chemicals that contribute to pain in the human body.

Our BA predictions for acetaminophen on SCN9A and SCN10A are -9.60 kcal/mol and -9.29 kcal/mol, respectively, indicating that acetaminophen is a good binder on SCN9A. Furthermore, the predicted BA value on hERG from our model is -7.39 kcal/mol, which is higher than the hERG side effect threshold of -8.18 kcal/mol, validating the safety profile of acetaminophen on hERG.

Our predictions suggest that acetaminophen exhibits the highest inhibitory effect on the LATS2 protein, with a predicted BA value of -11.2 kcal/mol. LATS2 is a protein kinase that plays a significant role in cell growth regulation, apoptosis, and tumor suppression. It is associated with various diseases, including breast cancer, lung cancer, ovarian cancer, neurofibromatosis type 2 (NF2), and cardiovascular diseases. Inhibiting the LATS2 protein could lead to serious side effects, which might explain the potential reasons for the high side effects of acetaminophen, such as liver damage, allergic reactions, skin reactions, gastrointestinal issues, blood disorders, and kidney problems.    

Nonsteroidal anti-inflammatory drugs (NSAIDs), such as ibuprofen (Advil, Motrin), and naproxen (Aleve), are commonly used for the treatment of mild to moderate pain accompanied by swelling and inflammation. These medications can inhibit certain enzymes in the human body that are released due to tissue damage. Ibuprofen, a non-selective inhibitor of the enzyme cyclooxygenase (COX), plays a crucial role in the synthesis of prostaglandins through the arachidonic acid pathway. COX facilitates the conversion of arachidonic acid to prostaglandin H2 (PGH2) in the body, which is further transformed into other prostaglandins. By inhibiting COX, ibuprofen reduces the production of prostaglandins in the body, resulting in pain relief.

The predicted BA values of ibuprofen for SCN9A and SCN10A are -9.11 and -9.72 kcal/mol, respectively, indicating strong potency of ibuprofen on SCN10A. The predicted BA value for hERG is -7.13 kcal/mol, suggesting a safe hERG-blockade profile. Additionally, ibuprofen is predicted to be a potent inhibitor of LATS2, USP9X, and MTOR, which are the top three proteins with the largest absolute predicted BA values (-11.17, -10.68, -10.46 kcal/mol). Furthermore, the predicted BA value of ibuprofen on TRPM8 is -10.04 kcal/mol, validating its strong binding affinity to TRPM8, a thermosensitive ion channel implicated in pain signaling, particularly in cold-induced pain or cold allodynia.

Despite its effectiveness, ibuprofen can cause a number of side effects, including nausea, constipation or diarrhea, and indigestion (dyspepsia).

Naproxen, like other NSAIDs such as ibuprofen, inhibits COX, leading to analgesic and anti-inflammatory effects. It is also a potent inhibitor of sodium channels, as validated by the predicted BA values of -9.02 and -9.6 kcal/mol for SCN9A and SCN10A, respectively. The predicted BA value of -6.55 kcal/mol for hERG confirms the safety profile of naproxen on hERG. Our predictions indicate that naproxen may have side effects on other targets, with the top three predicted BA values being -11.35, -11.32, and -11.13 kcal/mol for CSNK2A2, FGFR2, and LATS2, respectively. This aligns with the known fact that naproxen can cause a range of potential side effects, including dizziness, headache, bruising, allergic reactions, and stomach pain \cite{maniar2018lowering}. Additionally, naproxen demonstrates strong inhibition of TRPM8 with a predicted BA value of -9.97 kcal/mol.

Opioids are powerful pain-relieving medications commonly prescribed for moderate to severe pain. Examples of opioid medications include oxycodone (OxyContin, Roxicodone), hydrocodone (Vicodin, Hysingla ER), fentanyl (Actiq, Fentora), and morphine (MS Contin), among others. They function by binding to opioid receptors in the brain, spinal cord, and other parts of the body, thereby reducing the perception of pain. Due to their potential for misuse, addiction, and overdose, these medications are subject to strict prescribing guidelines.

Oxycodone, a strong semi-synthetic opioid, is used medically to treat moderate to severe pain. Its mechanism of action involves interacting with opioid receptors in the central nervous system. The predicted BA values of oxycodone for SCN9A and SCN10A are -9.75 and -10.62 kcal/mol, respectively. The predicted BA value for hERG is remarkably low at -7.8 kcal/mol, indicating a low potential for hERG side effects. Oxycodone demonstrates strong binding potency to the top three proteins: ROS1, CSNK2A2, and OPRM1, with the largest predicted BA values being -11.77, -11.47, and -11.45 kcal/mol, respectively. Additionally, our predictions suggest that oxycodone can inhibit the TRPA1 (Transient Receptor Potential Ankyrin 1) protein, with a predicted BA value of -10.09 kcal/mol. TRPA1 is a thermosensitive ion channel involved in the detection and transmission of pain signals. It is known for its role in mediating various types of pain, particularly in response to chemical irritants and inflammatory stimuli.

Hydrocodone is indicated for the relief of acute pain, sometimes in combination with acetaminophen or ibuprofen. It is also used for the symptomatic treatment of the common cold and allergic rhinitis, often in combination with decongestants, antihistamines, and expectorants. Hydrocodone inhibits pain signaling in both the spinal cord and brain. Its actions in the brain can also lead to euphoria, respiratory depression, and sedation \cite{trescot2008opioid}.

In our predictions, hydrocodone demonstrates good binding affinities for SCN9A and SCN10A, with BA values of -9.72 and -10.56 kcal/mol, respectively. The predicted BA value for hERG is -8.16 kcal/mol, suggesting a low potential for side effects on hERG. Hydrocodone has the potential to cause serious side effects on the top three proteins: ROS1, CSNK2A2, and TACR1, with predicted BA values of -11.98, -11.40, and -11.36 kcal/mol, respectively. Additionally, our findings indicate that hydrocodone is a strong binder to the TRPA1 protein, with a predicted BA value of -9.94 kcal/mol.

Some medications prescribed to manage depression and prevent epileptic seizures have been found to relieve chronic pain. Tricyclic antidepressants used in the treatment of chronic pain include amitriptyline and nortriptyline (Pamelor). Anti-seizure medications used for chronic nerve pain include gabapentin (Gralise, Neurontin, Horizant) and pregabalin (Lyrica).

Amitriptyline, a tricyclic antidepressant, has been used for decades to treat depression and has been investigated for its analgesic properties in pain-related conditions \cite{bryson1996amitriptyline}. Our predicted BA values for SCN9A and SCN10A are -9.74 and -10.04 kcal/mol, respectively, validating the potency of amitriptyline in pain treatment according to our predictions. The predicted BA value of amitriptyline on hERG is -8.25 kcal/mol, indicating a potential side effect on hERG.

The three strongest predicted BA values are for LATS2, HRH1, and KCNA3 proteins, with values of -11.08, -11.01, and -10.61 kcal/mol, respectively. Gabapentin, a structural analogue of the inhibitory neurotransmitter gamma-aminobutyric acid (GABA), was originally developed as an anti-epileptic medication. It is now widely used to treat neuropathic pain \cite{kukkar2013implications}. Our predictions suggest that gabapentin has the potential to inhibit SCN9A and SCN10A, with BA values of -9.0 and -9.35 kcal/mol, respectively. Moreover, gabapentin is predicted to have no side effects on hERG, with a BA value of -6.85 kcal/mol. In addition, our predictions show that the three strongest predicted BA values are for LATS2, KCNA3, and FGFR2, with values of -10.94, -10.61, and -10.6 kcal/mol, respectively.

\subsection{Nearly optimal lead compounds from screening and repurposing}

We dedicate our efforts to finding more potential inhibitors of the two pain targets, SCN9A and SCN10A, through the screening and repurposing processes in this section. In the process of screening and repurposing, we utilized 110 ML models to predict the cross-target binding affinity. In addition to considering potency, we also ensured that the optimal ranges for the ADMET properties and SAS (as listed in Table.\ref{table_1}), as well as the hERG side effect, were all well satisfied. SCN9A and SCN10A are not only major pain targets but also key pharmacological targets in pain treatment. To identify more promising potent compounds for these two targets, we utilized the 110 inhibitor datasets as a source of inhibitor compounds.

During the screening process, we selected potent inhibitor compounds with experimental BA values below -9.54 kcal/mol from the inhibitor datasets of the two pain targets, SCN9A and SCN10A. We then evaluated a series of other properties. It's important to note that if a designated inhibitor of one target demonstrates high efficacy on the other target, it is not considered a side effect. This is because it is common for an inhibitor to be potent on both major pain targets simultaneously. However, we still need to evaluate the potential for side effects on the other 108 protein targets, as well as hERG. We require predicted BA values greater than -9.54 kcal/mol to exclude side effects, except for hERG, which has a stricter requirement of BA values greater than -8.18 kcal/mol.

For repurposing, we assess the binding potency of all weak inhibitors in the other 108 datasets on the two pain targets, SCN9A and SCN10A. Therefore, we select inhibitors with experimental BA values greater than -9.54 kcal/mol and identify those with predicted BA values less than -9.54 kcal/mol on the two pain targets. In our search for inhibitors with repurposing potential on the pain targets, these inhibitors should have no side effects on the other 107 proteins, as well as hERG. Furthermore, we also study the optimal range of ADMET properties and synthetic accessibility.

\begin{figure}[!tpb]
\centering
\includegraphics[width=13cm]{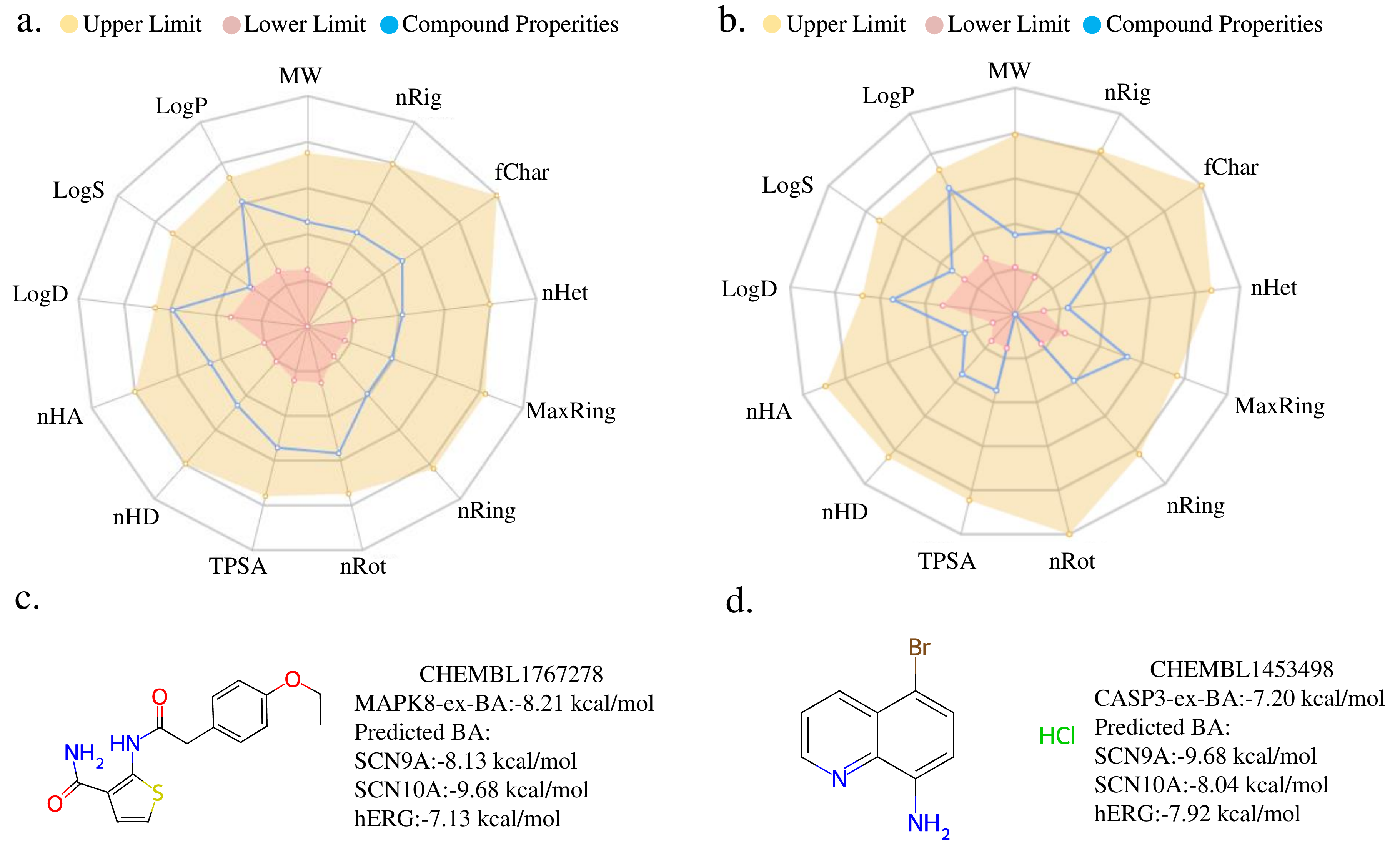}
\caption{Assessment of 13 ADMET properties for those molecular compounds with repurposing potentials. ${\bf a}$ and ${\bf b}$ indicate the evaluations of ADMET properties of two compounds CHEMBL1767278 and CHEMBL1453498, and ${\bf c}$ and ${\bf d}$ represent their chemical graphs and predictions of side effects, respectively. The boundaries of yellow and red regimes in ${\bf a}$ and ${\bf b}$ show the upper and lower limits of the optimal ranges for 13 ADMET properties, respectively. The blue curves suggest values of the specified 13 ADMET properties. The details of these property abbreviations are as following: MW (Molecular Weight), logP (log of octanol/water partition coefficient), logS (log of the aqueous solubility), logD (logP at physiological pH 7.4), nHA (Number of hydrogen bond acceptors), nHD (Number of hydrogen bond donors), TPSA (Topological polar surface area), nRot (Number of rotatable bonds), nRing (Number of rings), MaxRing (Number of atoms in the biggest ring), nHet (Number of heteroatoms), fChar (Formal charge), and nRig (Number of rigid bonds). }
\label{radarmap}
\end{figure}

It is not easy to find inhibitors that satisfy all the aforementioned requirements. In the end, we identified two inhibitor compounds, CHEMBL 1767278 from the MAPK8 dataset and CHEMBL 1453498 from the CASP3 dataset, for repurposing. The former is predicted to have BA values of -8.13 and -9.68 kcal/mol on SCN9A and SCN10A, respectively, while the latter is predicted to have values of -9.68 and -8.04 kcal/mol, indicating their potency on SCN10A and SCN9A, respectively. Their predicted BA values on hERG are -7.13 and -7.92 kcal/mol, respectively, suggesting favorable side effect profiles. The representations of the two compounds and their side effect predictions are provided in Fig.\ref{radarmap}c and d, respectively. Furthermore, these two compounds are predicted to have no binding or side effects on the remaining 96 and 99 proteins, respectively. We also evaluated additional ADMET properties of these two molecular compounds using the ADMETlab 2.0 prediction solver (https://admetmesh.scbdd.com/). Fig.\ref{radarmap}a and b show that the two compounds fall within the optimal ranges of these ADMET properties. For more details on the meaning and optimal ranges of the 13 ADMET properties, please refer to Table S4 in the Supporting Information.

\begin{figure}[!tpb]
\centering
\includegraphics[width=14cm]{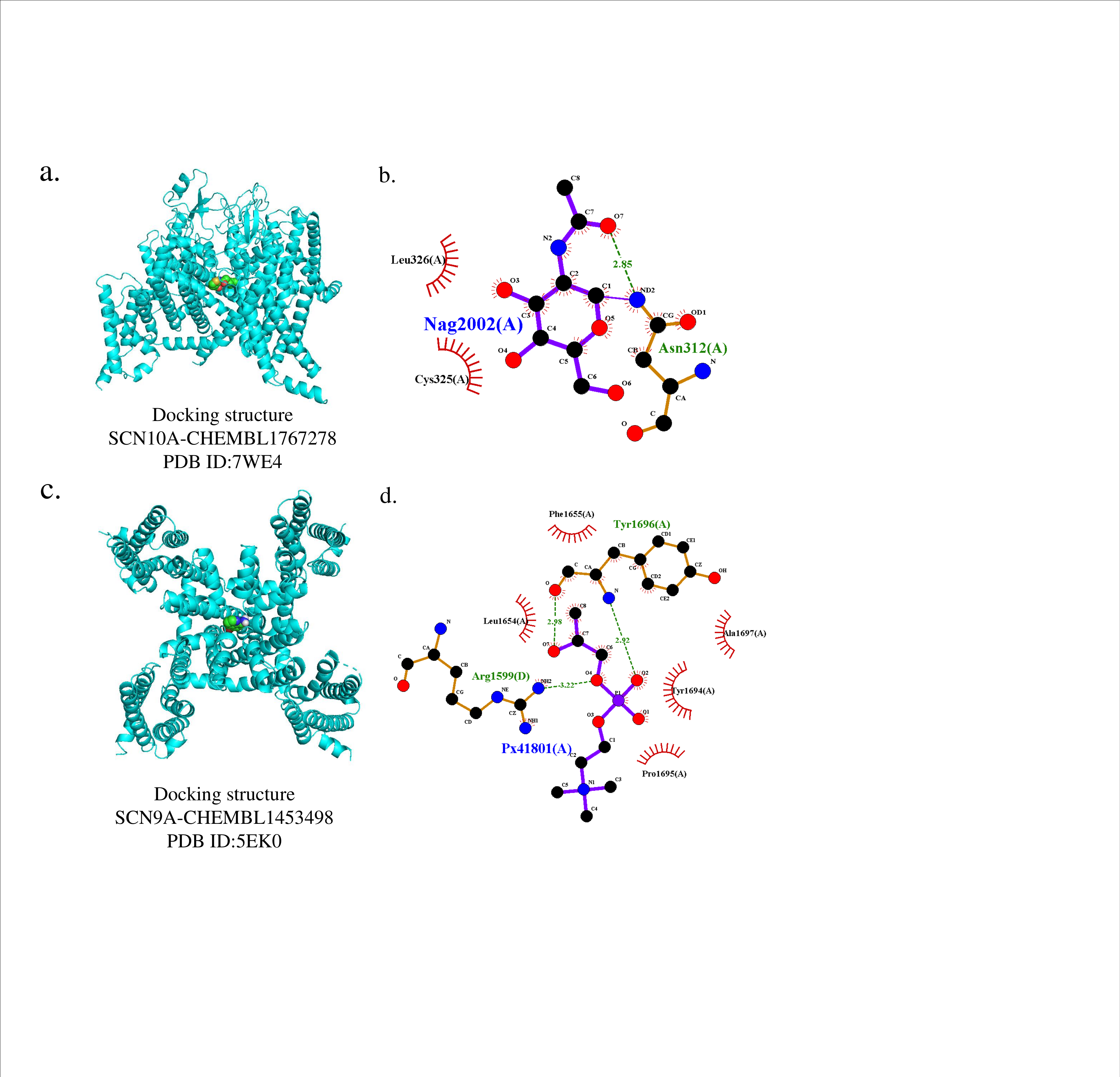}
\caption{The docking structure of our two optimal lead compounds bound to two pain targets SCN0A and SCN10A, and their 2D interaction diagrams. We use AutoDock Vina to implement the protein-ligand docking, and find the hydrogen bonds generated during the docking of two compounds.}
\label{docking}
\end{figure}

Next, we investigated the molecular interactions between the two inhibitors and the two main pain targets, SCN9A and SCN10A, using the software AutoDock Vina \cite{huey2012using}. Fig.\ref{docking}a, c shows the 3D protein-ligand docking structures, and Fig.\ref{docking}b, d shows the 2D interaction diagrams of the two compounds, CHEMBL1767278 and CHEMBL1453498, respectively. Due to the structural complexity of SCN9A and SCN10A, we focused on the docking between the inhibitors and the central sites of the targets. AutoDock Vina generated 9 docking poses with different docking scores calculated from its scoring function. In our figures, we selected the pose with the highest affinity (kcal/mol), where hydrogen bonds are formed between the inhibitors and the two pain targets SCN9A and SCN10A. In the docking of compound CHEMBL1767278 (see Fig.\ref{docking}b), one strong hydrogen bond with Asn312 (2.85 Å) is formed, while in the docking of compound CHEMBL1453498 (see Fig.\ref{docking}d), three hydrogen bonds with Tyr1696 (2.98 Å, 2.92 Å) and Arg1599 (3.22 Å) are formed. The predicted binding energies of these two compounds with SCN10A and SCN9A are both -9.68 kcal/mol. Additionally, we found that neither of the two compounds formed a covalent bond with the side chains of the targets during the docking process, suggesting that hydrogen bonds play vital roles in the interaction between the atoms.

\section{Methods}

\subsection{Datasets}

All inhibitor datasets were collected from the Chembl database (https://www.ebi.ac.uk/chembl/) for all proteins in the present DTI network, which was informed by four investigated sodium channels (Na$\text{v}$1.3, Na$\text{v}$1.7, Na$\text{v}$1.8, and Na$\text{v}$1.9, corresponding to encoded genes SCN3A, SCN9A, SCN10A, and SCN11A, respectively). Since the predictive results of machine learning-based models depend on high-quality and quantity of data, we set the minimal size of the collected inhibitor datasets to be 250 and obtained a total of 111 datasets, including SCN9A and SCN10A. The datasets for SCN3A and SCN11A were not included due to their small data size. The labels for these datasets are binding affinities (BAs) obtained using the formulas BA = 1.3633*log${10}$$K_i$ and $K_i$=IC${50}$/2 \cite{kalliokoski2013comparability}. As hERG is a key target for side effects in virtual screening of drug design, an inhibitor dataset was also collected from the Chembl database. All details of the datasets are provided in Table S3 in the Supporting Information.

\subsection{Molecular fingerprints}
Molecular fingerprints represent the property profiles of a molecule, typically in the form of vectors where each element represents the presence, degree, or frequency of a specific structural characteristic. These fingerprints can be used as features in machine learning (ML) models. The original molecular fingerprints for the inhibitors in the collected 111 datasets are 2D SMILES strings. In this study, we utilized two types of latent-vector molecular fingerprints in the ML models: bidirectional encoder transformer fingerprint (BET-FP) and autoencoder fingerprint (AE-FP). These fingerprints were generated from pre-trained models based on natural language processing (NLP) algorithms such as transformers and sequence-to-sequence autoencoders \cite{chen2021extracting, winter2019learning}. They are latent embedding vectors with a length of 512, obtained by encoding the 2D SMILES strings of the inhibitor compounds using the pre-trained models.

\subsubsection{Sequence-to-sequence auto-encoder fingerprint}
Recently, Winter et al. proposed a data-driven unsupervised learning model for extracting molecular information embedded in the SMILES representation \cite{winter2019learning}. Their approach involved using a sequence-to-sequence autoencoder to translate one form of molecular embedding to another by capturing the chemical structure's complete description in the latent space between the encoder and decoder. This translation model was capable of extracting physical and chemical information during the embedding process, enabling the translation to a distinct molecular representation with the same semantics but different syntax. Notably, the translation model was trained on a large dataset of chemical structures and could be used to extract molecular fingerprints for query compounds without the need for retraining or labels.

Typically, the translation model consists of encoder and decoder networks. The encoder network compresses the essential information from the input SMILES, which is then fed as input to the decoder network. Convolutional neural networks (CNNs) and recurrent neural networks (RNNs) were employed in the decoder, with fully connected layers mapping the output of the CNN or concatenated cell states of the RNN to intermediate vector embeddings between the encoder and decoder networks. Consequently, the decoder incorporates RNN networks with latent vectors as input. To extract more physical and chemical information from the latent vectors, the translation model was extended based on a classification model that predicts molecular properties using these vectors. The output of the RNN in the decoder network represents the probability distributions of various characters in the translated molecular embeddings. During the training of the autoencoder model, the loss function consists of the sum of cross-entropies between the predicted probability distributions and the correct characters encoded in a one-hot format, as well as the mean squared errors of the molecular property predictions made by the classification model.

In this study, the translation model was trained on approximately 72 million molecular compounds obtained from the ZINC and PubChem databases. The compounds underwent preprocessing, including filtering based on criteria such as molecular weight, number of heavy atoms, partition coefficient, and more. By training the translation model on this processed dataset, the resulting model generated embedding vectors that served as molecular fingerprints.

\subsubsection{Bidirectional transformer }

Recently, Chen et al. developed a deep learning network that was pretrained on millions of
unlabeled molecules using a self-supervised learning (SSL) platform to extract predictive molecular fingerprints \cite{chen2021extracting}. The SSL approach employed the bidirectional encoder transformer (BET) model, which relies on the attention mechanism. Unlike constructing a complete encoder-decoder framework, SSL utilized the decoder network solely for encoding the molecular SMILES.

In the SSL pre-training platform, the input consisted of molecular SMILES strings. Pairs of real SMILES and masked SMILES were created by hiding a certain number of meaningful symbols within the strings. The model was then trained using these data-mask pairs in a supervised manner with the SSL method. During the pretraining process, the masked symbols were learned by studying the unprocessed symbols in the SMILES, enhancing the understanding of the SMILES language. Data masking was performed as a preprocessing step before training the model with SSL. A total of 51 symbols were considered as elements in the SMILES strings. The SMILES were used as input to train the model, with a maximum length set to 256. Two special symbols, '$<s>$' and '$<\backslash s>$', were added to the beginning and end of the SMILES strings. If a string's length was less than 256, the '$<pad>$' symbol was used to complete the SMILES string. For the data masking process, 15\% of the symbols in the SMILES were manipulated, with 80\% being masked, 10\% remaining unchanged, and the remaining 10\% randomly changed.

The BET module plays a crucial role in achieving SSL from a substantial number of SMILES strings. It utilizes the attention mechanism in the transformer module to extract the importance of each symbol in the SMILES sequence. The BET module consists of eight bidirectional encoder layers, where each layer includes a multi-head self-attention layer and a subsequent fully connected feed-forward neural network. Each self-attention layer has eight heads, and the embedding size of the fully connected feed-forward layers is 1024. During training, the Adam optimizer with a weight decay of 0.1 is employed, and the loss function chosen is cross-entropy. The input SMILES have a maximum length of 256, including the special symbols added at the two ends, and each symbol is embedded in a dimension of 512. Consequently, the resulting molecular embedding matrix consists of 256 embedding vectors, each with a dimension of 512.

The transformer module offers high parallelism capability and training efficiency, allowing for the use of a large amount of SMILES to train deep learning models. In this study, SMILES strings from the Chembl, PubChem, and ZINC databases, either individually or fused together, were used to train three separate pre-trained models. The resulting transformer-based molecular embeddings generated from the pre-trained models using the Chembl database were utilized as molecular fingerprints.

\subsection{Machine learning models}
Three classic machine learning algorithms, namely gradient boosting decision tree (GBDT), support vector machine (SVM), and random forest (RF), are employed to construct our ML models. The GBDT algorithm, an ensemble approach, possesses several advantages such as resistance to overfitting, insensitivity to hyperparameters, and ease of implementation. Consequently, it is competitive when training with small datasets and can yield better prediction performance compared to deep neural networks (DNNs) and other common ML algorithms. However, it is important to note that one of the challenges of GBDT is to strike a balance between accuracy and efficiency for large datasets. The algorithm assembles multiple weak learners (individual trees) into an iterative prediction model. While weak learners may produce suboptimal predictions individually, the combination of all weak learners through the ensemble approach helps reduce overall errors. The primary procedure of GBDT involves learning decision trees, where most of the time is consumed in finding the best split points. GBDT has already demonstrated good performance in various quantitative structure-activity relationship (QSAR) prediction tasks \cite{jiang2021ggl,jiang2020boosting}. In this study, the GBDT algorithm provided by the Scikit-learn library (version 0.24.1) was utilized.

Support Vector Machine (SVM), introduced by Cortes and Vapnik, is a non-probabilistic kernel-based supervised learning method that maps input vectors into a high-dimensional feature space \cite{cortes1995support}. The core concept behind SVM is to identify the optimal decision boundary that separates different classes in the feature space. This decision boundary is defined by a hyperplane that maximizes the margin between the support vectors and the data points closest to the decision boundary. SVM offers advantages such as high efficiency in high-dimensional spaces, robustness against overfitting, and versatility. However, SVM also has some limitations, including computational complexity and sensitivity to parameter tuning.

Random Forest (RF), developed by Breiman, is an ensemble of decision trees where the predictions of individual trees are averaged to obtain an ensemble performance \cite{breiman2001random}. It employs a bootstrap sampling technique, and each decision tree uses only a subset of randomly chosen samples and features, starting with a trunk that splits into multiple branches before reaching the leaves. The leaf nodes represent the final prediction, while all other nodes are assigned with molecular features. RF is widely used in solving QSAR prediction problems and often does not require a complex feature selection procedure. Moreover, it is robust to redundant features and exhibits insensitivity to parameter variations. 

We collected a total of 111 inhibitor datasets in our DTI network. The three aforementioned ML algorithms were used to build ML models for these datasets. The details of the hyperparameters for these three ML algorithms are provided in Table S5 in the Supporting Information. In the ML models, we used two types of molecular fingerprints, namely BET and AE fingerprints, to embed the inhibitor compounds. Our ML models were created by pairing these molecular fingerprints with the GBDT, SVM, or RF algorithm. Consequently, we built a total of 111 ML models, each corresponding to one inhibitor dataset.

For each dataset, six individual models were constructed by combining BET and AE fingerprints with the three ML algorithms. The average of the predictions from these six individual models was considered as our final binding affinity prediction, which we refer to as the consensus method for prediction. The consensus results typically outperform those obtained from individual models. We compared the prediction results using the three different algorithms and found that the SVM algorithm with the consensus method performed the best among the other algorithms using individual fingerprints. This was validated using a set of provided samples, as shown in Table S6 in the Supporting Information. To reduce the impact of randomness, each individual ML model was trained ten times using different random seeds, and the average of the ten predictions was considered as the final result for each individual model. Additionally, the Pearson correlation coefficients (R) and root mean square deviation (RMSD) of ten-fold cross-validations for the 111 datasets are presented in Table S7 of the Supporting Information.

\section{Conclusion}
Pain is a complex sensory and emotional experience that serves as a protective mechanism in response to potential or actual tissue damage. It can be categorized into different types, such as psychogenic pain, physical pain, and neuropathic pain, based on various factors. Physical pain occurs when there is actual or potential damage to tissues, such as injury or surgery. Nociceptors, specialized sensory receptors, detect noxious stimuli and transmit signals to the brain, resulting in the perception of pain. Neuropathic pain, on the other hand, originates from damage or dysfunction of the nervous system itself. It may be caused by conditions such as nerve compression, diabetes, or trauma, and is often described as shooting, burning, or electric shocks accompanied by abnormal sensations. Physical pain and neuropathic pain share common underlying neurological mechanisms. Sodium channels, particularly Na$\text{v}$1.3, Na$\text{v}$1.7, Na$\text{v}$1.8, and Na$\text{v}$1.9, play a significant role in the generation and transmission of pain signals in various pain conditions. Consequently, sodium channel blockers that specifically target these channels have been actively explored as potential therapeutic interventions for pain. By modulating the activity of sodium channels, it is possible to reduce abnormal pain signaling associated with different pain conditions. However, progress in drug design for pain treatment has been relatively slow, and there is a need for more treatment options to be investigated.

Sodium channels are attractive targets for the development of pain medications. Pain affects complex molecular and biological activities in the nervous system, involving significant protein-protein interactions (PPI) in different brain regions. The development of pain treatment medications must take into account the influence of drugs on the PPI networks of pain targets. In this study, we construct an extended drug-target interaction (DTI) network informed by four pain-related sodium channels. We develop a machine learning framework to screen and propose additional drug candidates for pain reduction. We utilize two molecular fingerprints generated by advanced natural language processing (NLP) models based on transformer and autoencoder algorithms. These fingerprints are then used to build predictive machine learning models employing three common machine learning algorithms: support vector machine (SVM), gradient boosting decision tree (GBDT), and random forest (RF). A consensus model combining the predictions from these algorithms is used to enhance the overall predictive performance. Additionally, we apply these machine learning models to reevaluate the side effects of existing pain-relieving medications. Our ML models are also employed to analyze the repurposing potential of existing inhibitor compounds on major pain targets and screen for possible side effects associated with these inhibitors. Furthermore, we implement the assessment of ADMET properties using machine learning predictions. Finally, we identify a group of promising compounds for major pain targets. Further testing through in vitro or animal experiments is necessary to evaluate the toxicity and blood-brain barrier permeability characteristics of these candidate compounds.

Our machine learning-based framework provides a novel method for searching candidate compounds for pain relief and can be generalized for other diseases with neurological implications. While the sodium channel genes studied in this work are associated with pain perception and pain disorders, it is important to note that pain is a complex and multifactorial phenomenon involving numerous other factors and pathways. Further research is needed to fully understand the roles of these sodium channels in pain processing and to explore their potential as therapeutic targets for pain management.

\section*{Data and code availability}
The related datasets studied in present work are available at: \url{ https: //weilab.math.msu.edu/DataLibrary/2D/}. Codes of the calculation of two molecular fingerprints are available via \url{https: //github.com/WeilabMSU/OUD-PPI}.

\section*{Acknowledgements}
This work was supported in part by NIH grants R01GM126189 and R01AI164266, NSF grants
DMS-2052983, DMS-1761320, and IIS-1900473, NASA grant 80NSSC21M0023, MSU Foundation, Bristol-Myers Squibb 65109, and Pfizer. The work of Jian Jiang and Bengong Zhang was supported by the National Natural Science Foundation of China under Grant No. 11971367, No.12271416, and No.11972266.


\section*{Competing interests}
The authors declare no competing interests.


\end{document}